\documentclass[preprint,showpacs,amsmath]{revtex4}

\usepackage{graphicx}
\usepackage{dcolumn}
\usepackage{bm}
\def\PLB{{\it Phys. Lett.} }
\def\PRL{{\it Phys. Rev. Lett.} }
\def\NP{{\it Nucl. Phys.} }
\def\NPPS{{\it Nucl. Phys. Proc. Suppl.} }
\def\PR{{\it Phys. Rev.} }

\def\JP{{\it J. Phys.} }

\def\ZPC{{\it Z. Phys. } }
\def\bee{\begin{eqnarray}}
\def\eee{\end{eqnarray}}
\def\nn{\nonumber\\}

\def\jpg1#1#2#3{J.Phys.{\bf G}: Nucl. Part. Phys. {\bf #1},#2(#3)}

\def\nb{\bar{\nu}}
\def\s#1{\not\!\!{#1}}
\begin{document}
\title
{\bf Polarization and Distribution Function of $\Lambda_b$ Baryon}
\author{Tsung-Wen Yeh}
\affiliation{Institute of Physics, National Chiao-Tung University, Hsinchu 300, Taiwan}
\begin{abstract}
The polarization of $\Lambda_b$ baryon has been measured in ALEPH, OPAL and DELPHI experiments. A significant loss on the transfer of the $b$ quark polarization to the $\Lambda_b$ baryon polarization has been noticed. This implies that the hadronization effects can not be neglected. Therefore, we may make use of the polarization measurements to look for suitable model for the $\Lambda_b$ distribution function. To investigate nonperturbative effects implied in the $\Lambda_b$ polarization, we construct four models based on a perturbative QCD factorization formula. The models are the quark model (QM), the modified quark model (MQM), the parton model (PM), and the modified parton model (MPM). The modified models mean the models having transverse degrees of freedom and the associated soft radiative corrections having been resummed. The quark and parton models can not describe all experiments in the same time. On the other hand, the modified models can have the power to explain all data in the same formalism. 
\end{abstract}
\maketitle
\section{Introduction}
The polarization of bottom baryons $\Lambda_b$'s has been measured by ALEPH \cite{ALEPH}, OPAL \cite{OPAL} and DELPHI \cite{DELPHI}. The ALEPH data showed that the $\Lambda_b$ polarization has value ${\rm P}=-0.23^{+0.24}_{-0.20}(\makebox{stat.})\pm 0.08(\makebox{syst.})$. The OPAL data indicated the polarization ${\rm P}=-0.56^{+0.20}_{-0.13}(\makebox{stat.})\pm 0.09(\makebox{syst.})$. The DELPHI experiment gave ${\rm P}=-0.49^{+0.32}_{-0.30}(\makebox{stat.})\pm 0.17(\makebox{syst.})$. Although these three measurements are compatible with each other, the $\Lambda_b$ polarization still has a wide range of value from $+0.01$ to $-0.79$. To improve the situation, it is better to find out a sensitive measurable quantity on the polarization. However, this is very difficult before we can have a more qualitative understanding on the spin properties of $\Lambda_b$ baryons. This paper is intended  to understand the behind mechanisms by constructing physical models based on perturbative QCD formalism. 

Measurement of a large longitudinal polarization of the $\Lambda_b$ may indicate  the polarization of primary $b$ quark produced from a $Z^0$ decay. The $b$ quarks produced in the reaction $e^+ e^-\to Z^0\to b \bar{b}$ are highly polarized with polarization ${\rm P}=-0.94$ \cite{CLOSE,Mannel-Schular,KPTung}. The corrections from hard gluon emissions and mass effects can change the polarization of the final state $b$ quarks by only $3\%$ \cite{KRZ,JW}. The $b$ quark can fragment into mesons and baryons. The decays of b mesons into spin zero pseudoscalar states do not retain any polarization information. The hadronization to $b$ baryons might preserve a large fraction of the initial $b$ quark polarization. In the heavy quark mass limit, the spin degrees of freedom of $b$ quark are decoupled from a spin-zero light diquark. The initial polarization of $b$ quark can therefore be preserved until the $\Lambda_b$ decays. The higher mass $b$ baryon states can decay into the $\Lambda_b$ baryon but transfer little spin degrees of freedom. These effects have been estimated from different scenarios as about $30\%$. This leads to that the final $\Lambda_b$ polarization could be ${\rm P}=-0.6 \sim 0.70$ \cite{FP,JK}.     
   
The ALEPH Collaboration measured the $\Lambda_b$ polarization by employing the method suggested by Bovicini and Randall \cite{BR}. In the ratio $y({\rm P})=\langle E_\ell({\rm P})\rangle/\langle E_{\bar{\nu}}({\rm P})\rangle$ with $\langle E_\ell\rangle$ and $\langle E_{\bar\nu}\rangle$ the averaged lepton and  antineutrino energies in the laboratory reference frame, the fragmentation effects are largely cancelled out. Also, the spectra of the electrons and anti-neutrinos produced from the inclusive semileptonic decays of polarized $\Lambda_b$ baryons are harder relative to the spectra of unpolarized decays. The ALEPH Collaboration proposed to measure the ratio ${\rm R}({\rm P})=y({\rm P})/y(0)$, which is a Lorentz invariant quantity. The $\Lambda_b$ polarization is then extracted from a comparison between the measured ratio ${\rm R}^{\makebox{\tiny ALEPH}}=1.12\pm 0.10$ and a Monte Carlo simulation ratio ${\rm R}^{\tiny\rm MC}({\rm P})$ with varying $\rm P$. Because the $\Lambda_b$ polarization is best defined in the rest frame of the $\Lambda_b$ baryons, one can rewrite the ratio $\rm R$ in terms of the averaged variables in the rest frame to determine the polarization
\bee\label{PR}
{\rm P}=\frac{\langle E_\ell^{\ast} \rangle\langle E_{\bar{\nu}}^{\ast} \rangle(1-{\rm R})}{\langle E_\ell^{\ast}\rangle\langle P_{\bar{\nu}}^{\ast} \rangle {\rm R}-\langle E_{\bar{\nu}}^{\ast}\rangle \langle P_\ell^{\ast} \rangle}\ , 
\eee
where the star variables denote the averaged quantities in the rest frame and are calculated with ${\rm P}=-1$. The above equation can be investigated by different theoretical models. For example, if we apply the free quark model to calculate the star variables, we can obtain the polarization $\rm P=-0.23$, which is closed to the ALEPH's result \cite{ALEPH}.  The situation will become interesting as we apply the same model for the DELPHI experiment. The DELPHI experiment measured the same ratio ${\rm R}^{\makebox{\tiny DELPHI}}=1.21^{+0.16}_{-0.14}$ and obtained $\rm P^{\makebox{\tiny DELPHI}}=-0.49^{+0.32}_{-0.30}$. In the same way, it is easy to check that substituting DELPHI's ratio ${\rm R}^{\makebox{\tiny DELPHI}}$ into Eq.~(\ref{PR}) can derive a different value $\rm P=-0.38$ in free quark model. On the other hand, in the same model, if we employ the OPAL's polarization ${\rm P}^{\makebox{\tiny OPAL}}=-0.56$ into the ALEPH's and DELPHI's Monte Carlo simulation ratios $\rm R^{\makebox{\tiny MC}}$, we can extract the central value of corresponding $\rm R$ as ${\rm R}_1=1.30$ and $\rm R_2=1.27$, respectively. This seems to imply that there requires more investigations to find a consistent picture for $\Lambda_b$ polarization. That is we need to find a model which can explain the experiments self-consistently. The model dependence in the equation for $\rm P$, such as Eq.~(\ref{PR}), arises from the $z$ variable $z_{\ell(\bar{\nu})}=\langle P^{\ast}_{\ell(\bar{\nu})}\rangle/\langle E^{\ast}_{\ell(\bar{\nu})}\rangle$. Using the $z$ variables, Eq.~(\ref{PR}) can be recast as
\bee
{\rm P}=\frac{(1-{\rm R})}{z_{\bar{\nu}}{\rm R}-z_\ell}\ .
\eee 
It will become clear in the following sections that different models would give different values of ratio $z_{\ell}$ but almost the same $z_{\nb}$ due to the characteristics of the lepton and anti-neutrino spectra. 

In order to explore the mechanisms controlling the spin properties of polarized $\Lambda_b$ baryons, we shall investigate four models. They are the (free) quark model (QM), the modified quark model (MQM), the parton model (PM), and the modified parton model (MPM). The parton model describes the probability of finding the $b$ quark carrying a momentum fraction of the momentum of the $\Lambda_b$ baryon by a parton distribution function. The quark model assumes that the $\Lambda_b$ baryon contains only one $b$ quark and two light quarks and the corresponding parton distribution function is just a delta function of the momentum fraction. This means that the $b$ quark carries almost all the $\Lambda_b$ baryon momentum. The modified quark model and the modified parton model mean that the quark model and the parton model contain additional Sudakov form factor and transverse momentum. The Sudakov form factor arises from a resummation over radiative corrections of soft gluons and have the effects to enhance the  perturbative QCD contributions.

We emphasize the importance of transverse degrees of freedom of partons
inside a $\Lambda_b$ baryon in our analysis. First, the transverse momenta regularize the divergences when the outgoing $c$ quark in the process $b\to c W$ is approaching the end point. Second, the transverse momenta also enhance the contributions from the spin vector along the polarization direction.   
For completeness, we also introduce the intrinsic transverse momentum for the distribution function. We assume that the form of the intrinsic transverse momentum part of the parton distribution function can be parameterized as $\exp{[-tM^2b^2]}$ with an impact parameter $b$, which is the conjugate variable of the transverse momentum. The other factors are the $\Lambda_b$ baryon mass $M$ and a dimensionless parameter $t$. The impact parameter $b$ will be integrated out in our PQCD formalism. The $z$ variables are functions of $t$. To determine the parameter $t$ we rely on the experiments. The OPAL Collaboration determined the polarization by comparing the measured distribution of the ratio $E_{\bar\nu}/E_\ell$ against a simulation of this ratio using the JETSET Monte Carlo event generator. The polarization $\rm P^{\makebox{\tiny OPAL}}$ of OPAL experiment and results of the ALEPH and DELPHI experiments will determine the range of parameter $t$.

The arrangement of our paper is as follows. In the next section,
we shall demonstrate the factorization formula for the inclusive semileptonic decay of the $\Lambda_b$ baryon. In this formula, the hard scattering amplitude, describing the short distance subprocess $b\to c \ell \bar{\nu}$, convolutes with a jet function and an universal soft function. For simplicity, we shall assume that the charm quark mass can be ignored. That is we shall neglect the corrections like $m_c^2/M^2$ with $m_c$ the charm quark mass. This approximation is less than $10\%$ and is safe as compared with the accuracy of the experiments. However, it requires to consider the collinear divergences due to our ignorance of the charm quark mass. The jet function is then necessary for absorbing the collinear divergences. The universal soft function involves the $b$ quark matrix element. The matrix element contains a large scale factor, the $b$ quark mass $M_b$. To have a well established matrix element, we need to employ heavy quark effective theory (HQET) to scale out this large scale. We also need to separate the leading order matrix elements in $1/M_b$ expansion from the higher order ones. We shall develop a description for separating the leading order from the higher order mass corrections. This description is equivalent to the OPE approach. In Section 3, we shall construct four models based on the factorization formula. Section 4 gives the numerical result. The conclusion is given in Section 5. 

\section{Factorization Formula }
We shall investigate the quadruple differential decay rate for polarized $\Lambda_b$ baryons, ${\Lambda_b}\to X_c\ell\nb$,
\bee\label{dr}
\frac{d^4\Gamma}{dE_\ell dq^2 dq_0 d\cos\theta_\ell}
&=&\frac{|V_{cb}|^2{G_F^2}}{256\pi^4M}L^{\mu\nu}W_{\mu\nu}\;\;, 
\eee
where $M$ denotes the mass of the $\Lambda_b$ baryon, $L^{\mu\nu}$ represents the leptonic tensor  
\bee\label{lc}
 L^{\mu\nu}=Tr[\s{P}_\ell \Gamma^{\mu} \s{P}_{\nb} \Gamma^{\mu}]\;\;,
\eee
and $W_{\mu\nu}$ means the hadronic tensor  
\bee\label{hc}
W_{\mu\nu}&=&(2\pi)^3\delta^{(4)}(P_{\Lambda_b}-q-P_{X_c}){\mathop{\sum}_{X}}
\langle \Lambda_b |\bar{c}\Gamma_{\mu}^{\dag} b| X_c\rangle
\langle X_c |\bar{c}\Gamma_{\nu}b|\Lambda_b\rangle
\frac{d^3{\bf P_{X_c}}}{(2\pi)^3 2 E_{X_c}} \;\; 
\eee
where $\Gamma_{\mu}$ denotes the $V-A$ operator $\gamma_{\mu}(1-\gamma_5)$, $|X_c\rangle$ means the hadronic states containing a charm quark and $q$ is total momentum carried by the lepton and anti-neutrino. We choose the normalization for $\Lambda_b$ state $|\Lambda_b\rangle$ as $\langle\Lambda_b(P^\prime_{\Lambda_b},S)|\Lambda_b(P_{\Lambda_b},S)\rangle =(2\pi)^3(2P^0_{\Lambda_b})\delta^3(\vec{S}-\vec{S}^\prime)\delta^3(\vec{P}_{\Lambda_b}-\vec{P}^\prime_{\Lambda_b})$. The kinematical variables $E_\ell$, $q$, $q_0$ and $\cos\theta_\ell$ are expressed as follows. We choose the $\Lambda_b$ baryon rest frame such that the initial $\Lambda_b$ baryon momentum, $P_{\Lambda_b}$, and the final state lepton and anti-neutrino momenta, $p_\ell$ and $p_{\bar{\nu}}$, can be defined as 
\bee
P_{\Lambda_b}=\frac{M}{\sqrt{2}}(1,1,{\bf 0}), p_\ell=(p_\ell^+,0,{\bf 0}), 
p_{\nb}=(p_{\nb}^+,p_{\nb}^-,{\bf p}_{\nb\perp})\;.
\eee
The variables $E_\ell$, $q$, and $q_0$ are related to $p_\ell^+, p_{\nb}^+, p_{\nb}^- $ as $E_\ell=p_\ell^+/\sqrt{2}$, 
$q^2=2p_\ell^+ p_{\nb}^-$, and $q_0=(p_\ell^+ + p_{\nb}^+ + p_{\nb}^-)/\sqrt{2}$. We let $P_b=P_{\Lambda_b}-l$ represent the $b$ quark momentum whose square is set as $P_b^2\approx M_b^2$ with $M_b$ the $b$ quark mass. The momentum $l$ of the light degree of freedom of the $\Lambda_b$ baryon can have a large plus component and small transverse components ${\bf l}_{\perp}$. The final state charm quark momentum is equal to $P_c=P_{\Lambda_b}-l-q$. The angle $\theta_\ell$ is defined as the angle between the third component of $p_\ell$ and that of the $b$ quark polarization vector, $S_b$. The differential decay rate can also be rewritten in terms of $E_{\bar{\nu}}, y, y_0$ and $\theta_{\bar{\nu}}$ with $E_{\bar{\nu}}$ the anti-neutrino energy and corresponding angle $\theta_{\bar{\nu}}$. Because the right hand side of Eq.~(\ref{dr}) is independent of which parameterization for the leptonic variables, we shall use both parameterizations for the differential decay rate. 

It is convenient to use the scaled variables $x_{\ell(\bar{\nu})}=2E_{\ell({\bar{\nu}})}/M$, $y=q^2/M^2$,
and $y_0=2q_0/M$. The integration regions for $x_{\ell(\bar{\nu})}$, $y$ and $y_0$ are 
\bee
0\le x_{\ell(\bar{\nu})}\le 1, \hspace{1cm} 0\le y \le x_{\ell(\bar{\nu})},\hspace{1cm} \frac{y}{x_{\ell(\bar{\nu})}}+x_{\ell(\bar{\nu})}\le y_0\le 1+y.
\eee
 Note that we have chosen $M$ as the scale factor.  For simplicity, we have chosen $m_c=0$ and left $m_c\ne 0$ to other work.  This approximation is safe as compared with the accuracy of available experiments. The leading corrections to this approximation is of order $O(m_c^2/M_b^2)$ being less than $10\%$.

By optical theorem, the hadronic tensor $W^{\mu\nu}$ can be related to the forward matrix element $T^{\mu\nu}$ through the formula 
\bee
W^{\mu\nu}=-\frac{\mbox{Im}(T^{\mu\nu})}{\pi}.
\eee
The lowest order of $T^{\mu\nu}$ is defined as
\bee
T^{\mu\nu}(P_{\Lambda_b},q,S)&=&-i\int d^4 y e^{iq\cdot y}
\langle \Lambda_b(P_{\Lambda_b},S)|{\cal T}[
J^{\dag\mu}(0),J^{\nu}(y)]|\Lambda_b(P_{\Lambda_b},S)\rangle 
\eee
with $J^{\mu}=\bar{q}\gamma^{\mu}(1-\gamma_5)b$ the V $-$ A current. The forward matrix element can be expressed in the momentum space
\bee 
T^{\mu\nu}(P_{\Lambda_b},q,S)&=&-i\int \frac{d^4P_b}{(2\pi)^4} \makebox{Tr}[S^{\mu\nu}(P_b-q)T(P_{\Lambda_b},S,P_b)],
\label{st}
\eee
where the trace is taken over the fermion indices and color indices, the hard function $S^{\mu\nu}(P_b-q)$ describes the short distance  
decay subprocess, $b\to W c$, and the soft function $T(P_{\Lambda_b},S,P_b)$ denotes the long distance matrix element  
\bee\label{t1}
T(P_{\Lambda_b},S,P_b)=\int d^4 y e^{iP_b\cdot y}\langle \Lambda_b(P_{\Lambda_b},S)|\bar{b}(0)b(y)
|\Lambda_b(P_{\Lambda_b},S)\rangle.  
\eee
Because we are only interested in the leading contributions in this note, we are required to separate the leading contributions from subleading contributions. To specify the leading contributions, we also need to consider correction terms. The correction contributions may come from radiative correction terms like $\alpha_s^n$ and power correction terms like $1/M^m$ for $n, m\ge 1$.  Among the radiative corrections, the contributions from soft gluons have logarithms $\alpha_s\log(m^2/Q^2)$ will become dominate at the end points, at which the final state quark is approaching on-shell. As discussed in the Introduction, the hard gluon emissions can only contribution about $3\%$. Therefore, we shall retain the soft gluon contributions at the end points. To discuss the power correction terms, we need to be more careful.  As investigated in the OPE and HQET approach, the power corrections can have two sources one from the short distance expansion for the forward matrix element and the other one from the heavy quark mass expansion for the expanded matrix elements. Here, we shall present a different approach in which the leading order matrix elements are in terms of nonlocal heavy quark currents composed of heavy quark effective fields in HQET. 

We now demonstrate this description. To start up, we express the forward matrix element 
\bee\label{fwme2}
iT^{\mu\nu}(P_{\Lambda_b},q,S)&=&\int \frac{d^4P_b}{(2\pi)^4} \makebox{Tr}[S^{\mu\nu}(P_b-q)T(P_{\Lambda_b},S,P_b)]\nn &+&\int \frac{d^4P_b}{(2\pi)^4} \int \frac{d^4 k^\prime}{(2\pi)^4} \makebox{Tr}[S^{\mu\nu}_{\alpha}(P_b-q,k^\prime)T^\alpha(P_{\Lambda_b},S,P_b,k^\prime)]+\cdots
\eee
by including a higher order term from triple parton matrix elements containing gluon fields
\bee
&&T^\alpha(P_{\Lambda_b},S,P_b,k^\prime)\nn
&=&\int d^4 y \int d^4 z e^{iP_b\cdot y} e^{i(P_b-k^\prime)\cdot z}\langle \Lambda_b(P_{\Lambda_b},S)|\bar{b}(0)(-ig A^\alpha(z))b(y)
|\Lambda_b(P_{\Lambda_b},S)\rangle  
\eee
with $A^\alpha(z)$ the gluon fields. We shall employ the light-cone gauge $A^+=n\cdot A=0$. To continue, it is useful to introduce the light-cone vectors $p$ and $n$ in the $+$  and $-$ directions, respectively. These two vectors satisfy properties $p^2=n^2=0$ and $p\cdot n=P_{\Lambda_b}\cdot n$. The $\Lambda_b$ baryon momentum $P_{\Lambda_b}$ is then recast as  
\bee
P^{\mu}_{\Lambda_b} &=& p^{\mu} + \frac{M^2}{2 p\cdot n}n^{\mu}\ . 
\eee
For the $b$ quark inside the $\Lambda_b$ baryon, we parameterize its momentum $P_b$ as
\bee
P_b^{\mu}&=& z {p}^{\mu} 
+ \frac{P_b^2 + P_{b\perp}^2}{2 P_b\cdot n }
 n^{\mu} + P_{b\perp}^{\mu} \\ 
&=& \hat{P}_b^{\mu} + \frac{P_b^2 - M_{b}^2}{2 P_b\cdot n }
n^{\mu}, 
\eee  
where $\hat{P}_b^2=M_b^2$ is the on-shell part of $P_b$ and the momentum fraction $z$ defined by $z=P^+_b/P^+_{\Lambda_b}=1-l^+/P^+_{\Lambda_b}$. By the parameterization of $P_b$, the $b$ quark propagator is then expressed as 
\bee
\frac{i}{\s{P}_b -M_b + i\epsilon}=\frac{i(\hat{\s{P}}_b + M_b)}
{P^2_b -M^2_b + i\epsilon}+ \frac{i\s{n}}{2 P_b\cdot n +i\epsilon}\;\;.
\label{heavyprop}
\eee
Since we allow $l^+\approx P^+_{\Lambda_b}$, therefore we have $z\sim \Lambda/M$ with $\Lambda$ about $\Lambda_{\makebox{\tiny QCD}}$ and $M_b\sim M$. The second term of Eq.~(\ref{heavyprop}) having power like $\frac{1}{zM}\sim \frac{1}{\Lambda}$ is as large as the first term. We thus do not need to take into account the power correction from the collinear part of the $b$ quark propagator. This demonstration can be better understood from the fact that the $b$ quark is almost on shell and has a large quark mass. Therefore, there is no collinear divergences associated with the $b$ quark.  The remaining work is to separate the leading terms of the hard function $S^{\mu\nu}$ from the higher order terms. The hard function $S^{\mu\nu}$ is a function of $l$. We can make Taylor expansion for $S^{\mu\nu}$ with respect to $l^+$. This is because the momentum $l$ has a large plus component $l^+=\xi p$ with $\xi=1-z$. By performing Taylor expansions for $S^{\mu\nu}(l)$ and $S^{\mu\nu}_\alpha(l,k^\prime)$ around $S^{\mu\nu}(\xi p)$ and $S^{\mu\nu}_\alpha(\xi p,\xi^\prime p)$, we then obtain 
\bee\label{smunu}
S^{\mu\nu}(l)&=&S^{\mu\nu}(\xi p)+\frac{\partial S^{\mu\nu}}{\partial l^{\alpha}}|_{l=\xi p}(l-\xi p)^\alpha+\cdots\ ,\nn
S^{\mu\nu}_\alpha(l,k^\prime)&=&S^{\mu\nu}_\alpha(\xi p,\xi^\prime p)+\frac{\partial S^{\mu\nu}_\alpha}{\partial l^{\alpha}}|_{l=\xi p,k^\prime=\xi^\prime p}(l-\xi p)^\alpha+\frac{\partial S^{\mu\nu}_\alpha}{\partial k^{\prime\alpha}}|_{l=\xi p,k^\prime=\xi^\prime p}(k^\prime-\xi^\prime p)^\alpha\cdots\ .
\eee 
The Ward identity ensures the following equation to hold
\bee
\frac{\partial S^{\mu\nu}}{\partial l^{\alpha}}(\xi p)=S^{\mu\nu}_\alpha(\xi p, \xi p)\ .
\eee 
The contributions from $S^{\mu\nu}_\alpha$ terms are power suppressed than $S^{\mu\nu}(\xi p)$ by at least $O(1/M^2)$. The effects of the second terms in Eq.~(\ref{smunu}) are to replace the gluonic field operators in the second term in Eq.~(\ref{fwme2}) by covariant derivative operators. Let's explain this. By adding the second terms from Eqs.~(\ref{smunu}) and (\ref{fwme2}), respectively, we have
\bee
\int \frac{d^4 l}{(2\pi)^4}S^{\mu\nu}_\alpha(\xi p,\xi p)(l-\xi p)^\alpha T(l)+\int \frac{d^4 l}{(2\pi)^4}\int \frac{d^4 k^\prime}{(2\pi)^4}S^{\mu\nu}_\alpha(\xi p,\xi^\prime p)T^{\alpha}(l,k^\prime)\ .
\eee
In light cone gauge $n\cdot A=0$, we can rewrite the above equation as
\bee\label{cladd1}
\int \frac{d^4 l}{(2\pi)^4}\int \frac{d^4 k^\prime}{(2\pi)^4}S^{\mu\nu}_\alpha(\xi p,\xi^\prime p)w^{\alpha}_{\alpha^\prime}[l^{\alpha\prime} T(l)(2\pi)^4\delta^4(k^\prime)+ T^{\alpha^\prime}(l,k^\prime)]
\eee
where the projection tensor $w^{\alpha}_{\alpha^\prime}=g^{\alpha}_{\alpha^\prime}-p^{\alpha}n_{\alpha^\prime}$ has been employed. Using the identity
\bee
(2\pi)^4\delta^4(k^\prime)=\int d^4 z e^{i k^\prime\cdot(y- z)}\ ,
\eee
the bracket in Eq.~(\ref{cladd1}) can be recast as
\bee
\int dy^4 \int d^4z e^{il\cdot y}e^{i k^\prime\cdot(y- z)}\langle \Lambda_b(P_{\Lambda_b},S)|\bar{b}(0)D^\alpha(z)b(y)
|\Lambda_b(P_{\Lambda_b},S)\rangle  
\eee
with $D^{\alpha}=i\partial^\alpha-ig A^\alpha$.

The $b$ quark field in the leading matrix element $T$ still contains a large phase $\exp{(-iM_b v\cdot x)}$ with $v$ the $\Lambda_b$ velocity. This is unsuitable to define a matrix element at low energies. To solve this, we can employ the heavy quark effective theory (HQET). In HQET, We can rescale the $b$ quark field, $b(x)$, into $b_v(x)=\exp{(iM_b v\cdot x)}\frac{1+\s{v}}{2}b(x)$. The rescaled $b_v$ field is a small fluctuation quantity of coordinate, since the remaining scale in its phase factor is only about $\Lambda_{QCD}$ scale. In HQET, $P_b$ is parameterized as $P_b=M_b v +k$, with $k$ the residual momentum. The rescaled $b$ quark field, $b_v(x)$, carries the residual momentum $k$ and has a small effective mass $\bar{\Lambda}$, with $\bar{\Lambda}\equiv M-M_b$. Under the heavy quark mass expansion  
\bee
b_v(x)=h_v(x)+O(\frac{1}{M})+\cdots\ ,
\eee
the matrix element $T$ in terms of $b_v$ can be expanded as
\bee
T=T_0+O(\frac{1}{M^2})+\cdots\ .
\eee 
The $T_0$ is in terms of an universal effective heavy field $h_v$, which is defined as the $b_v$ field in the infinite mass limit $M_b\to\infty$. The missing of $O(\frac{1}{M})$ term is due to the equation of motion. The expression for $T_0$ is easily written down as
\bee
T_0=\int d^4 y e^{ik\cdot y}\langle \Lambda_b(v,S)|\bar{h}_v(0)h_v(y)
|\Lambda_b(v,S)\rangle\ .  
\eee
Note that we have replaced the hadronic state vector $|\Lambda_b(P_{\Lambda_b},S)\rangle$ by its equivalent representation $|\Lambda_b(v,S)\rangle$. The normalization of $|\Lambda_b(v,S)\rangle$ is large than the usual normalization by a factor $M^{\frac{1}{2}}$. We shall skip the derivation on how to derive the above equation. To derive leading order contributions, we still need to extract the leading spin structure of $T_0$. This can be achieved by means of Fierz identity. As a result, the leading order forward matrix element $T^{\mu\nu}$ takes the form 
\bee\label{exp1}
T^{\mu\nu}(P_{\Lambda_b},q,S)
&\approx& -i\int \frac{d^4 k}{(2\pi)^4} \left\{ 
[S^{\mu\nu}(k=\xi p,q)\s{P}_b][T_0(P_{\Lambda_b},S,k=\xi p)\frac{\s{n}}{4P_b\cdot n}]\right. \nn
&&\hspace{1cm}\left. -[S^{\mu\nu}(k=\xi p,q)\s{S}_b\gamma_5]
[T_0(P_{\Lambda_b},S,k=\xi p)\frac{\s{n}\gamma_5}{4S_b\cdot n}]\right\}+O(\frac{1}{M^2})\;\;,
\eee
where we have inserted the Fierz identity
\bee
I_{ij}I_{mn}=\frac{1}{4}(\gamma^{\mu})_{im}(\gamma_{\mu})_{jn}+\frac{1}{4}(\gamma^{\mu}\gamma_5)_{im}(\gamma_5\gamma_{\mu})_{jn}+\cdots
\eee
where $i,j,m,n$ denote the fermion indices and the dots represent the other gamma matrix would result in higher order terms. 

We now briefly describe how to derive the factorization formula 
for the inclusive semileptonic decay $\Lambda_b\to X_q\ell\nu$. The details about the derivation of the following factorization formula can be found in \cite{LIYU1}. We shall only demonstrate the main ideas and not try to give a repeated proof. The formula for the quadruple differential decay rate can be expressed as a convolution integral over the soft function $S$, the jet function $J$ and a hard function $H$
\begin{eqnarray}\label{as}
\frac{1}{\Gamma^{(0)}}\frac{d^3\Gamma}{dxdydy_0d\cos\theta}
&=&\frac{M^2}{2}\int^{z_{\rm max}}_{z_{\rm min}}{dz} \int d^2{\bf l_\bot}
\nonumber \\
&&\times
S(z,{\bf l_\bot},\mu)J(z,P_c^-,{\bf l_\bot},\mu)H(z,P_{c}^-,\mu)\;,
\end{eqnarray}
where $x=x_{\ell}(x_{\bar{\nu}})$ and $\theta=\theta_\ell(\theta_{\nb})$ and $\Gamma^{(0)}=\frac{G_F^2}{16\pi^3}|V_{qb}|^2{M}^5$. The scale $\mu$ is introduced as a renormalization and factorization scale. The transverse momentum ${\bf l}_\bot$ has been reintroduced for regularization of the end point singularities \cite{LIYU1}. The end point singularities arise from the end point region $x\to 1$ and $y, y_0\to 0$. The charm quark (assumed as massless) has a large minus component $P_c^-=(1-y/x)M/\sqrt{2}$ and a small plus component $P_c^+=(1-y_0-y/x)M/\sqrt{2}$. This implies there is a very small invariant $P_c^2=M^2(1-y_0+y)$, which leads to an on-shell jet subprocess. The ${\bf l}_\bot$ integrals can be finished only when we know the exact dependence of the jet function on ${\bf l}_\bot$. But the jet function is nonperturbative and cannot be determined theoretically, so far. Fortunately, this difficulty for integration over ${\bf l}_\bot$ can be removed by means of a Fourier transformation for the jet function into its impact space representation as
\bee
J(z,P_q^-,{\bf l_\bot},\mu)=\int \frac{d^2{\bf b}}{(2\pi)^2}{\tilde J}(z,P_c^-,{\bf b},\mu)e^{i{\bf l_\bot}\cdot \bf b}\ .
\eee
The $\bf l_\bot$ integrals then decouple from the jet function and the remaining factor $e^{i{\bf l_\bot}\cdot \bf b}$ is then associated with the soft function.
The factorization formula Eq.~(\ref{as}) can also be applied to the case with loop corrections. With the Fourier transformation for $\bf l_\bot$, the Feynman rule for the radiative gluon cross over the final state cut should be modified with an extra phase factor $e^{i{\bf l_\bot}\cdot \bf b}$. The upper and lower limits of $z$ are chosen as $z_{\max}=1$ and $z_{\min}=x$. The lower limit $z_{\min}=x$ is from the jet function. The upper limit $z_{\max}=1$ is chosen to fill the kinematical gap between $M_b$ and $M$. The Fourier transformation of Eq.~(\ref{as}) into the impact $\bf b$ space then takes the form   
\bee\label{asb}
\frac{1}{\Gamma^{(0)}}\frac{d^3\Gamma}{dxdydy_0d\cos\theta_l}
&=&\frac{M^2}{2}\int^{1}_x{dz} \int \frac{d^2{\bf b}}{(2\pi)^2}
\nonumber \\
&&\times{\tilde S}(z,{\bf b},\mu){\tilde J}(z,P_c^-,{\bf b},\mu)
H(z,P_{c}^-,\mu)\;.  
\eee

To deal with the collinear and soft divergences resulting from
the radiative corrections for massless parton inside the jet, 
the resummation technique is necessary and these divergences could be resummed into a Sudakov form factor \cite{LIYU1}. The jet function is then re-expressed into the form 
\bee
{\tilde J}(z,P_c^-,b,\mu)=\exp{[-2s(P_c^-,b)]}{\tilde J}(z,b,\mu),
\label{js}
\eee 
where $\exp{[-2s(P_c^-,b)]}$ is the Sudakov form factor. The RG invariant Sudakov exponent has the expression up to one loop accuracy
\bee
s(P_c^-,b)&=&\frac{A^{(1)}}{2\beta_1}\hat{q}\ln({\frac{\hat q}{\hat b}})+\frac{A^{(2)}}{4\beta_1^2}(\frac{\hat q}{\hat b}-1)-\frac{A^{(1)}}{2\beta_1}(\hat{q}-\hat{b})\nn
&&-\frac{A^{(1)}\beta_2}{4\beta_1^2}\hat{q}
\left[\frac{\ln{(2\hat{b})}+1}{\hat{b}}-\frac{\ln{(2\hat{q})}+1}{\hat{q}}\right]\nn 
&&-\left[\frac{A^{(2)}}{4\beta_1^2}-\frac{A^{(1)}}{4\beta_1}\ln(\frac{e^{2\gamma -1}}{2})\right]\ln(\frac{\hat{q}}{\hat{b}})\nn
&&+\frac{A^{(1)}\beta_2}{8\beta_1^3}[\ln^2(2\hat{q})-\ln^2(2\hat{b})]
\eee 
with the variables
\bee
\hat{q}=\ln(\frac{P_c^-}{\Lambda})\ ,\hat{b}=\ln(\frac{1}{b\Lambda})\ .
\eee 
We choose the QCD scale $\Lambda=\Lambda_{\makebox{\tiny QCD}}$ to have the value $0.2$ GeV in the numerical analysis in section 4. The other factors are defined as 
\bee
\beta_1&=&\frac{33-2n_f}{12}\ ,\nn
\beta_2&=&\frac{153-19n_f}{24}\ ,\nn
A^{(1)}&=&\frac{3}{4}\ ,\nn
A^{(2)}&=&\frac{67}{9}-\frac{\pi^2}{3}-\frac{10}{27}n_f+\frac{8}{3}\beta_1\ln(\frac{e^\gamma}{2})\ .
\eee
The scale invariance of the differential decay rate in Eq.~(\ref{asb})
and the Sudakov form factor in Eq.~(\ref{js})
requires the functions $\tilde J$, $\tilde S$, and $H$ to obey the following
RG equations:
\begin{eqnarray}
{\cal D}{\tilde J}(b,\mu)&=& -2\gamma_q {\tilde J}(b,\mu)\;,
\nonumber \\
{\cal D}{\tilde S}(b,\mu)&=& -\gamma_S{\tilde S}(b,\mu)\;,
\nonumber \\
{\cal D}H(P_c^-,\mu)&=& (2\gamma_q+\gamma_S)H(P_c^-,\mu)\;,
\label{ter}
\end{eqnarray}
with
\begin{equation}
{\cal D}=\mu\frac{\partial}{\partial\mu}+\beta(g)
\frac{\partial}{\partial g}\;.
\end{equation}
$\gamma_q=-\alpha_s/\pi$ is the quark anomalous dimension in axial
gauge, and $\gamma_S=-(\alpha_s/\pi)C_F$ is the anomalous dimension of
$\tilde S$.
After solving Eq.~(\ref{ter}), we obtain the evolution of all the convolution
factors in Eq.~(\ref{asb}),
\begin{eqnarray}
{\tilde J}(z,P_q^-,b,\mu)&=& {\rm exp}\left[-2s(P_q^-,b)-2\int^\mu_{1/b}
\frac{d{\bar\mu}}{{\bar\mu}}\gamma_q(\alpha_s(\bar\mu))\right]
{\tilde J}(z,b,1/b)\;,
\nonumber\\
{\tilde S}(z,b,\mu)&=& {\rm exp}\left[-\int^\mu_{1/b}
\frac{d{\bar\mu}}{{\bar\mu}}\gamma_S(\alpha_s(\bar\mu))\right]f(z,b,1/b)\;,
\nonumber\\
H(z,P_c^-,\mu)&=& {\rm exp}\left[-\int^{P_c^-}_\mu
\frac{d{\bar\mu}}{{\bar\mu}}[2\gamma_q(\alpha_s(\bar\mu))
+\gamma_S(\alpha_s(\bar\mu))]\right]H(z,P_c^-,P_c^-)\;.
\label{un}
\end{eqnarray}
In the above solutions, we set the $1/b$ as an IR cut-off for
single logarithm evolution. 

For the initial soft function $f(z,b,1/b)$, we shall keep the intrinsic $b$ dependence by $f(z,b,1/b)\approx f(z,b)$. The $b$ dependence in $f(z,b)$ can support us a way to explore its effect in determining the polarization. We assume $f(z,b)$ to have the form 
\bee
f(z,b)=f(z)e^{-\Sigma(b)}\ ,  
\eee
and take an ansaze for parameterizing $\exp{[-\Sigma(b)]}$ as  
\bee
e^{-\Sigma(b)}=e^{-tM^2b^2}
\eee  
with an unknown parameter $t$. To avoid double counting for the contributions from transverse degrees of freedom, we need some modifications for the factorization formula. For the end point regime where the Sudakov suppression dominates, we employ the approximation 
\bee
f(z,b)=f(z)
\label{suda}
\eee
, while for other regimes which are not under the control of the
Sudakov suppression, we take into account the intrinsic $b$ dependence of $f(z,b)$ 
\bee
f(z,b)=e^{-tM^2b^2}f(z)\;\;.  
\label{renor}
\eee  
We make further approximations such that ${\tilde J}(z,b,1/b)={\tilde J}^{(0)}(z,b)$, and $H(z,P_c^-,P_c^-)=H^{(0)}(z,P_c^-)$. 

Combining the above results, we arrive at factorization formula as
\begin{eqnarray}
\frac{1}{\Gamma^{(0)}}\frac{d^4\Gamma}{dxdydy_0d\cos\theta}
&=&M^2\int^{1}_{x}
dz\int_0^{\infty}\frac{bdb}{4\pi}{\tilde J}^{(0)}(z,b)
H^{(0)}(z,P_{c}^-) e^{-S(P_c^-,b)} 
\nonumber \\
&&\times
\left\{
\begin{array}{cc}
f(z)& \makebox{for $x$ in end point regimes}\;, \\
\exp[-tM^2b^2]f(z)& \makebox{for $x$ in other regimes}\;,
\end{array}
\right.
\label{as1}
\end{eqnarray}
where 
\bee
S(P^-_c,b)=2 s(P^-_c,b)-\int^{P^-_c}_{1/b}\frac{d\bar{\mu}}{\bar{\mu}}
[2\gamma_q(\bar{\mu})+\gamma_S(\bar{\mu})]\;.
\eee
The parameter $t$ will be determined by experiment. From practical calculations, we find that the above difference between the distribution function with and without intrinsic transverse momentum contributions is very small. Therefore, we shall include the factor $\exp[-tM^2b^2]$ for entire range of $x$ in the numerical analysis.

Let's now discuss how to parameterize $T_0(k)$ defined in Eq.~(\ref{t1}).
As discussed in previous paragraphs that, at leading order, $T_0(k)$ is expanded in the form 
\bee
T_0(k)=\frac{1}{4}P_b\cdot n \s{p} f(z) 
- \frac{1}{4} S_b\cdot n \s{p}\gamma_5 g(z)\ . 
\eee
The unpolarized and polarized distribution functions, $f(z)$ and $g(z)$, are 
defined as
\bee
f(z)&=&\int \frac{d\lambda}{2\pi}
e^{i(1-z)\lambda n} 
\langle v,S|\bar{h}_v(0)\s{n}h_v(\frac{\lambda n}{P\cdot n})|v,S\rangle
\eee
and
\bee
g(z)&=&\int\frac{d\lambda}{2\pi}
e^{i(1-z)\lambda n} 
\langle v,S|\bar{h}_v(0)\s{n}\gamma_5 h_v(\frac{\lambda n}{P\cdot n})
|v,S\rangle\;. 
\eee
It is easy to show that $f(z)$ and $g(z)$, in the heavy quark limit,
share a common matrix element which could be described by an universal
distribution function, $f_{\Lambda_b}(z)$. This just reflects the heavy
quark spin symmetry.   
We adopt the distribution function proposed in \cite{LIYU1} in the form 
\bee
f_{\Lambda_b}(z)=\frac{N z^2(1-z)^2}{((z-a)^2 + \epsilon z)^2}\theta(1-z)\;. 
\eee 
The parameters $N$, $a$ and $\epsilon$ are fixed by first 
three moments of $f_{\Lambda_b}(z)$
\bee\label{eq31}
& &\int_0^1 f_{\Lambda_b}(z) dz=1\;,
\nn
& &\int^1_0 dz (1-z) f_{\Lambda_b}(z)={\bar \Lambda}/M +{\cal O}
(\Lambda^2_{\rm QCD}/M^2)\;,
\nn 
& &\int^1_0 dz (1-z)^2 f_{\Lambda_b}(z)=\frac{\bar \Lambda^2}{M^2} +\frac{2}{3}
K_b+{\cal O} (\Lambda^3_{\rm QCD}/M^3)\;,
\eee
where $\bar \Lambda=M-M_b$ and $K_b$ is to parameterize the matrix elemnet 
\bee
K_b=-\frac{1}{2M}\langle \Lambda_b|\bar{h}_v(0)\frac{(iD)^2}{2M^2}h_v(0)|\Lambda_b\rangle\ .
\eee 
By substituting the inputs 
\bee
M=5.641 {\rm GeV}\;,\;\;\;\;M_b=4.776 {\rm GeV}\;,\;\;\;\; K_b=0.012\pm
0.0026\;,
\label{con}
\eee
into Eq.~(\ref{eq31}), we determine the parameters $N$, $a$ and $\epsilon$ to be 
\begin{equation}
N=0.10615\;,\;\;\;\; a=1\;,\;\;\;\; \epsilon=0.00413\;.
\label{eq55}
\end{equation}
For simplicity we shall omit the subscript of $f_{\Lambda_b}(z)$ 
in the following text.

\section{Differential Decay Rates} 

In this section, we construct four models based on the factorization formula 
Eq.~(\ref{as1}). The models are the quark model (QM), the modified quark model (MQM), the parton model (PM), and the modified parton model (MPM). The charged lepton and anti-neutrino spectra for the decay $\Lambda_b\to X_q \ell\nb$ in the quark model are expressed as 
\bee\label{eq53}
\frac{1}{\Gamma^{(0)}}\frac{d^2 \Gamma^{\rm T}_{\rm QM}}{dxd\cos\theta}
&=&\left\{\begin{array}{l l}
\frac{x_\ell^2}{6}[(3-2x_\ell)-P\cos\theta_\ell(1-2x_\ell)] & \makebox{for $\ell$}\\
\frac{x_{\nb}^2}{6}x_{\nb} (1-x_{\nb})(1-P\cos\theta_{\bar{\nu}})& \makebox{for $\bar{\nu}$}
\end{array}\right.  
\eee
where $P$ and $\cos\theta_{\ell(\nb)}$ denote the polarization and cosine of the angle $\theta_{\ell(\nb)}$ between the third components of the lepton (anti-neutrino) momentum and the $\Lambda_b$ spin vector.

By taking into account Sudakov suppression from the resummation of
large radiative corrections, and substituting $f(z,b)=\delta (1-z)
\exp{[-t M^2b^2]}$,
$H^{(0)}=(x_\ell-y)[(y_0-x_\ell)-{\rm P}\cos\theta_\ell(y_0-x_\ell-2y/x_\ell)]$ and the Fourier transform of $J^{(0)}=\delta(P_c^2)$
with $P_c^2=M^2(1-y_0+y-p_{\bot}^2/M_B^2)$ into Eq.~(\ref{as1}),
we derive the lepton spectrum in modified quark model . The spectrum is,
after integrating Eq.~(\ref{as1}) over $z$ and $y_0$, described by
\bee\label{MQMu}
\frac{1}{{\Gamma}^{(0)}}\frac{d^2 \Gamma_{\rm MQM}}{dx_\ell d\cos\theta_\ell}&=&
M \int_0^{x}dy \int_0^{1/\Lambda}db e^{[-t^{\rm MQM} M^2 b^2]}
 e^{-S(P_q^-,b)}  (x_\ell-y) \eta\nn
&& \times\left[((1+y-x_\ell)-{\rm P\cos\theta_\ell}(1+y-x_\ell-2\frac{y}{x_\ell}) ) J_1(\eta M b)\right.\nn
&&\hspace{0.5cm}\left.-(\frac{2}{M b} \eta J_2(\eta M b) -\eta^2 J_3 (\eta M b))(1-{\rm P}\cos\theta_\ell)\right]\;,
\eee
where $P_c^-=(1-y/x_\ell)M/\sqrt{2}$, $\eta=\sqrt{(x_\ell-y)(1/x_\ell-1)}$ and
$J_1$,$J_2$,$J_3$ are the Bessel functions of order 1, 2 and 3, respectively.
Note that we have made an approximation by substituting $\exp{[-t M^2 b^2]}$ for the end point regimes. We also need the neutrino spectrum in modified quark model as
\bee\label{MQMnu}
\frac{1}{{\Gamma}^{(0)}}\frac{d^2 \Gamma_{\rm MQM}}{dx_{\nb} d\cos\theta_{\nb}}&=&
M \int_0^{x}dy \int_0^{1/\Lambda}db e^{[-t^{\rm MQM} M^2 b^2]}
 e^{-S(P_q^-,b)}  x_{\nb} (1-x_{\nb}) \eta^\prime J_1(\eta^\prime M b)\nn
&& \times(1-{\rm P\cos\theta_{\nb}}) 
\eee
with $\eta^\prime =\sqrt{(x_{\nb}-y)(1/x_{\nb}-1)}$.

The charged lepton spectrum in parton model is obtained
by adopting $H^{(0)}=(x_\ell-y)[(y_0-x_\ell-(1-z)y/x_\ell)-
{\rm P}\cos\theta_\ell(y_0-x_\ell-(1+z)y/x_\ell)]$ and
$P_c^2=M^2[1-y_0+y-(1-z)(1-y/x_\ell)]$. After integration over $y_0$, we then derive
\bee\label{e55a}
&&\frac{1}{{\Gamma}^{(0)}}\frac{d^2 \Gamma_{\rm PM}}{d x_\ell d\cos\theta_\ell}=\nn
&&\int_0^{x_\ell}dy \int_{x_\ell}^{1} dz f(z) (x_\ell-y)[(y+z-x_\ell)-{\rm P}
\cos\theta_\ell (y+z-x_\ell-2z\frac{y}{x_\ell})]\ .
\eee
In the same way, the neutrino spectrum can be written down
\bee\label{e55b}
\frac{1}{{\Gamma}^{(0)}}\frac{d^2 \Gamma_{\rm PM}}{d x_{\nb} d\cos\theta_{\nb}}=
\int_0^{x_{\nb}}dy \int_{x_{\nb}}^{1} dz f(z) x_{\nb}(z-x_{\nb})(1-{\rm P}\cos\theta_{\nb})\ .
\eee

The charged lepton spectra in the modified parton model takes into account  large perturbative corrections and nonperturbative intrinsic contributions with the expression as
\bee\label{MPMu}
\frac{1}{{\Gamma}^{(0)}}\frac{d^2 \Gamma_{\rm MPM}}{dx_\ell d\cos\theta_\ell}&=&
M \int_0^{x_\ell}dy \int^1_{x_\ell}dz \int_0^{1/\Lambda}db e^{[-t^{\rm MPM} M^2 b^2]}
 e^{-S(P_q^-,b)}  f(z)  (x_\ell-y) \eta\nn
&& \times\left[((z+y-x_\ell)-{\rm P}\cos\theta_{\ell}(z+y-x_\ell-2z\frac{y}{x_\ell}) ) J_1(\eta M b)\right.\nn
&&\hspace{0.5cm}\left.-(\frac{2}{M b} \eta J_2(\eta M b) -\eta^2 J_3 (\eta M b))(1-{\rm P}\cos\theta_\ell)\right]\;,
\eee
with $\eta=\sqrt{(x-y)(z/x_\ell-1)}$. The neutrino spectrum in modified parton model is also easily derived as
\begin{eqnarray}
\frac{1}{{\Gamma}^{(0)}}\frac{d^2 \Gamma_{\rm MPM}}{dx_{\nb} d\cos\theta_{\nb}}
&=&M \int_0^{x_{\nb}}dy \int^1_{x_{\nb}}dz \int_0^{1/\Lambda}db e^{-S(P_q^-,b)}
e^{-t^{\rm MPM} M^2 b^2} f(z)  x_{\nb}(z-x_{\nb}) \eta^\prime J_1(\eta^\prime M b)
\nn
&& \times(1-{\rm P}\cos\theta_{\nb})\ .
\end{eqnarray}
with $\eta^\prime=\sqrt{(x_{\nb}-y)(z/x_{\nb}-1)}$.

\section{Numerical Result}
The $\Lambda_b$'s produced in ALEPH, DELPHI and OPAL experiments are highly boosted in the laboratory frame. For the relativistic $\Lambda_b$'s, the forward-backward asymmetry of a decay product can be directly expressed in terms of a shift in the average value of its energy. The charged lepton also carried a residual sensitivity to the $\Lambda_b$ polarization. Because neither the $\Lambda_b$ four-momentum nor the lepton four-momentum can be fully reconstructed in the experiments, the ALEPH and DELPHI experiments proposed to measure the $\Lambda_b$ polarization, $\rm P$, through the variable $y$ suggested in \cite{BR} 
\bee
y=\frac{<E_\ell>}{<E_{\nb} >}\ .
\eee
However, there still exist many uncertainties suffered from experimental procedures on extracting the energy spectra. It requires normalizing the measured $y$ with an unpolarized simulated $y^{MC}(0)$. Therefore, the experimentally measured quantity is the ratio 
\bee
{\rm R}=\frac{y(\rm P)}{y^{MC}(0)} \;\;.
\eee 
ALEPH and DELPHI determined the polarization by comparing the measured value of ratio $\rm R$ from the Monte Carlo simulation ${\rm R}^{\makebox{\tiny MC}}({\rm P})$ with varying $\rm P$. The experimental results are $\rm R=1.12\pm 0.10$ and $\rm P=-0.23^{+0.24}_{-0.20}(\makebox{stat.})$ for ALEPH and $\rm R=1.21^{+0.16}_{-0.14}$ and $\rm P=-0.49^{+0.32}_{-0.30}(\makebox{stat.})$ for DELPHI, respectively. 
Theoretically, the $\Lambda_b$ polarization can be best defined in the rest frame. It is instructive to rewrite $y$ in terms of average variables in rest frame 
\bee\label{y}
y=\frac{\langle E_l^*\rangle +P\langle P_l^*({\rm P}=-1)\rangle}{\langle E_\nu^* \rangle+{\rm P}\langle P_\nu^*({\rm P}=-1)\rangle}\ ,
\eee
where the star average variables are evaluated with $\rm P=-1$. The average variables can be calculated from the formula
\bee
\langle a \rangle=\frac{\int  \int  a \frac{d^2 \Gamma}{dxd\cos\theta}dx d\cos\theta}{{\Gamma}^{(0)}}
\eee
by employing different models for the differential decay rate. It is much simplified in calculations of these averages, if the charged lepton and anti-neutrino average quantities are evaluated by their corresponding models for the differential decay rate. From these relations we can determine $\rm P$ in terms of $\rm R$ as
\bee\label{P}
{\rm P}=\frac{\langle E_l^* \rangle\langle E_\nu^* \rangle (1-{\rm R})}{\langle E_l^* \rangle \langle P_{\nb}^*\rangle {\rm R}-\langle E_{\nb}^* \rangle \langle P_l^* \rangle} \;.
\eee

We first compare the difference between the experimentally determined polarization $\rm P^{\rm EXP}$ and the theoretically evaluated polarization $\rm P^{\rm TH}$ in the four models QM, PM and MQM, MPM with parameter $t^{\rm MQM}=t^{\rm MPM}=0$. The result is shown in Table.~I. We can see that the theoretical polarizations are close to the ALEPH polarization $\rm P^{\makebox{\tiny ALEPH}}=-0.23$ but have a large deviation from the DELPHI polarization $\rm P^{\makebox{\tiny DELPHI}}=-0.49$. Among different model evaluations with one $\rm R$, their differences are very small. This implies that nonperturbative effects from distribution function over longitudinal momentum fraction and perturbative effects from Sudakov suppression are not important in determining the polarization.

We now turn on the parameters $t^{\rm MQM}$ and $t^{\rm MPM}$ to find out their values from experiments. It is interesting to note that the ratio $z_{l}=\langle P_l^* \rangle /\langle E_\ell^* \rangle$ is model dependent but the ratio $z_{\nb}=\langle P_{\nb}^* \rangle /\langle E_{\nb}^* \rangle$ is almost the same for all models. Using these two $z$ variables, we can rewrite Eq.~(\ref{P}) as
\bee\label{Pzl1}
{\rm P}=\frac{1-{\rm R}}{z_{\nb} {\rm R} - z_{\ell}}\ .
\eee
Since $z_{\nb}\approx 1/3$ for all models, we can further simplify the above equation into
\bee\label{Pzl2}
{\rm P}=\frac{3(1-{\rm R})}{{\rm R} - 3 z_{\ell}}\ .
\eee
We show the behaviors of $\rm P$ with respect to $z_\ell$ for ${\rm R}^{\makebox{\tiny ALEPH}}=1.12\pm 0.10$ (ALEPH) and $R^{\makebox{\tiny DELPHI}}=1.21^{+0.16}_{-0.14}$ (DELPHI) in Figs.~1 and 2, respectively. By applying the experimental bounds for $\rm P$, we can extract from Fig.~1 the $z_\ell$ range of $-\infty\le z_\ell\le -0.105$ for ALEPH and from Fig.~2 the range of $-1.75\le z_\ell\le -0.02$ for DELPHI. Theoretically, the $z_\ell$ range would be model dependent. Considering MQM and MPM and plotting the $z_\ell-t$ relation in Fig.~3, we can find that the maximum of $z_\ell$ can not be larger than $-0.05$ with $t\sim 0.3$. This is because the suppressions from the contributions of intrinsic transverse momentum are modeled by parameter $t$. By varying the value of $t$, we can easily change $z_\ell$. However, the fluctuations from the Bessel function in the differential decay rates would prevent the suppression of $t$ from becoming large. In the end, there exist a maximum bound for $z_\ell$. We notice that there is also a lower bound for $z_\ell$ as $z_\ell\ge -0.18$ with $t\sim 2$. We also hope that $t$ should be less than unity and close to zero to make the perturbative calculation reliable. Combining the above considerations, we can obtain the range for $z_\ell$ as $-0.12\le z_\ell \le -0.105$. The lower bound of $z_\ell$ comes from $t=0$. From Fig.~3, we can employ the $z_\ell$ range to derive the $t$ range. The $z_\ell-t$ relation is a bounce with maximum at $t\sim 0.3$ for both MQM and MPM. The $z_\ell$ range implies that $0\le t\le 0.1$. We have taken the convention that $t$ is located left to the maximum. This convention for choosing the value of $t$ can be further checked after we discuss the OPAL experiment. The $z_\ell-t$ plots for MQM and MPM are not much different. 

We now discuss the possible constraint over $z_\ell$ can be obtained from the OPAL experiment. The OPAL Collaboration employed a comparison between the measured $y3=E_{\nb}/E_{\ell}$ and the Monte Carlo simulated $y_3^{\makebox{\tiny MC}}$ to determine the polarization $P=-0.56^{+0.20}_{-0.13}(\makebox{stat.})$.  Applying the OPAL $\rm P$ to DELPHI and ALEPH experiments, we obtain $-0.6\le z_\ell \le -0.1$ for DELPHI and $-0.55\le z_\ell \le -0.105$ for ALEPH. The bound of $z_\ell$ is $-0.55\le z_\ell \le -0.105$. Looking at Fig.~2, one can see that the OPAL experiment give bound on $z_\ell$ in consistency with previous investigation. In summary, the ALEPH, OPAL and DELPHI experiments imply the range value of $0\le t\le 0.1$ and the corresponding range of $-0.12\le z_\ell\le -0.105$.

As a consistent check, we can write $\rm R$ in terms of $\rm P$ and $z_\ell$ as
\bee\label{Rzl}
R=\frac{3(1+{\rm P}z_\ell)}{(3+{\rm P})}\ .
\eee
By this equation, we can parameterize the Monte Carlo simulation ratios $R^{\makebox{\tiny MC}}({\rm P})$'s of ALEPH and DELPHI. We find that the value of $z_\ell\sim -0.075$ can be used for both experiments in a good approximation within $5-10\%$. In Fig.~4, we compare the $R-{\rm P}$ plots for $z_\ell\sim -0.075$ and $-0.12\le z_\ell\le -0.105$. The experimental bounds for ratio $\rm R$ can give constraints over $\rm P$. The combinnation of ALEPH and DELPHI experiments gives bound of $\rm P$ as $-0.79\le{\rm P}\le -0.05$, while the theoretical bounds are $-0.73\le{\rm P}\le -0.05$. The theoretical bounds for $\rm P$ being smaller than the experimental ones can be easily understood from the maximum bound of $z_\ell$ from theory is smaller than the $z_\ell$ employed in the Monte Carlo simulation performed by experiments. The difference between theory and experiment can be compensated by including higher order corrections, such as the mass corrections, etc.

\section{Conclusion}
 
In this paper we have constructed four models based on the PQCD factorization formula for $\Lambda_b \to X_c l \nb$. We used these models to investigate the physics implied by the ALEPH, OPAL and DELPHI experiments. We found that these experiments can be understood from theoretical models, the modified quark model and modified parton model. These two models contains intrinsic transverse momenta for partons, which are nonperturbative and parameterized by an exponential form with a parameter $t$. The parameter $t$ relates to the variable $z_\ell=\langle P_\ell^*({\rm P=-1})\rangle/\langle E_\ell^*\rangle$ with $\langle P_\ell^*({\rm P=-1})\rangle$ and $\langle E_\ell^*\rangle$ the average momentum and energy of charge lepton in the rest frame of $\Lambda_b$ baryon. We found that the ratio ${\rm R}=y({\rm P})/y(0)$ can be approximately expressed in terms of $\rm P$ and $z_\ell$. Using experimental results, we then determined the ranges of $z_\ell$ and $t$. 


\noindent
{\bf Acknowledgments:}
\hskip 0.6cm 

This work was supported in part by the National Science Council of R.O.C. under the Grant No. NSC89-2811-M-009-0024.

\newpage
Table.1 The values of the $\Lambda_b$ polarization 
 are predicted from the quark model, the parton model, the modified quark model 
and the modified parton model by employing the ALEPH and DELPHI experiments. The ALEPH and DELPHI experimental results are also shown for comparison. 
\vskip 1.0cm
\[\begin{array}{|lccc||c|c|c|} \hline\hline
{\rm P_{QM}} & {\rm P_{PM}} & {\rm P_{MQM}} & {\rm P_{MPM}}& {\rm P_{EXP}}& {\rm R} & {\rm Experiment}\\ \hline
-0.23^{+0.19}_{-0.17} & -0.23^{+0.19}_{-0.17} & -0.24^{+0.20}_{-0.17} &-0.24^{+0.20}_{-0.17}& -0.23^{+0.24}_{-0.20} & 1.12\pm 0.1 & {\rm ALEPH}\\\hline
-0.38^{+0.24}_{-0.24} & -0.38^{+0.24}_{-0.24} & -0.39^{+0.25}_{-0.24} &-0.39^{+0.25}_{-0.24}& -0.49^{+0.32}_{-0.30} & 1.21^{+0.16}_{-0.14} & {\rm DELPHI}\\
\hline\hline
\end{array} \]

\newpage
\begin{figure*}
\includegraphics{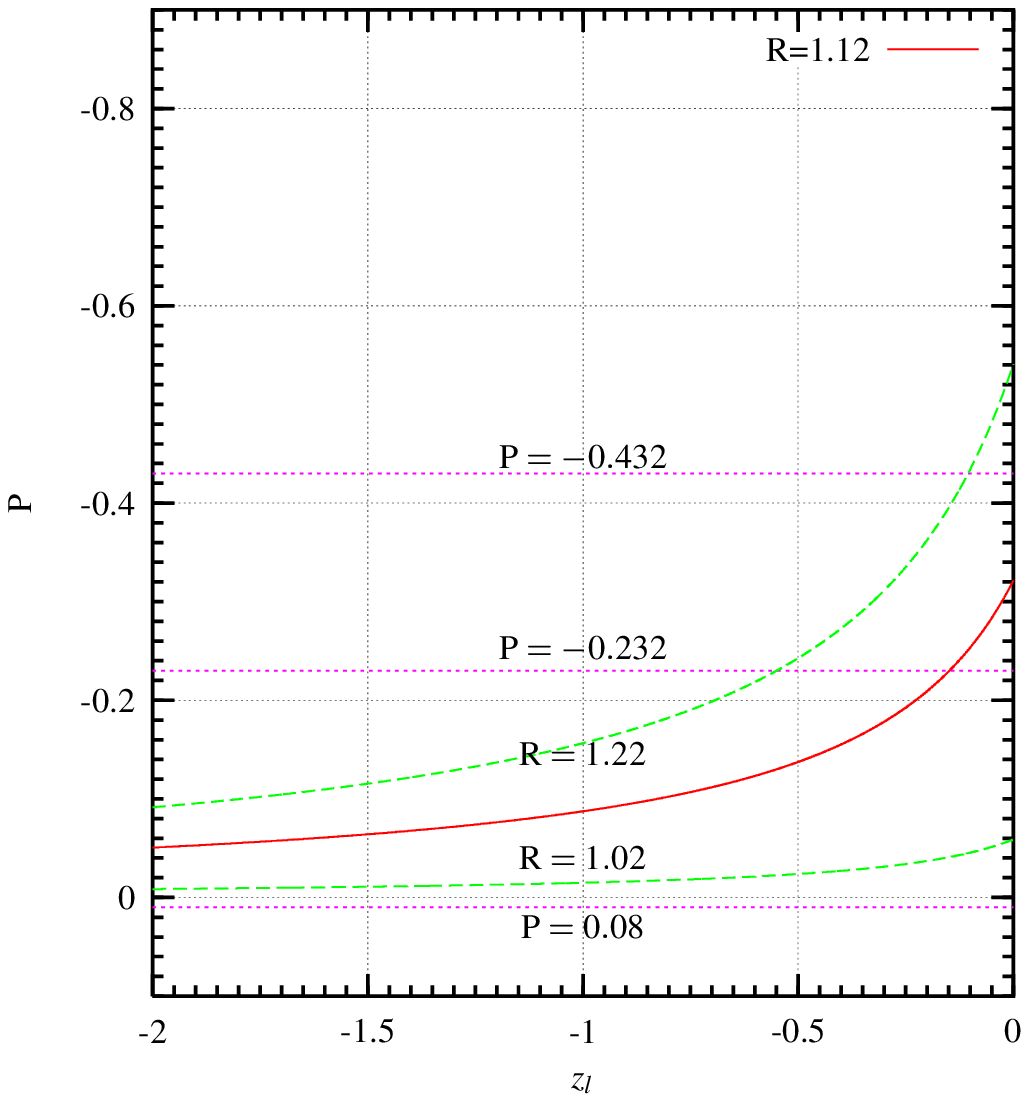}
\caption{ Plot of ${\rm P}$ vs. $z_{\ell}$ is shown by employing the ALEPH ratio ${\rm R}=1.12\pm 0.10$. The experimental polarization ${\rm P}={-0.23}^{+0.24}_{-0.20}$ is also shown to indicate the allowed range of $z_{\ell}$.}
\end{figure*}
\newpage
\begin{figure*}
\includegraphics{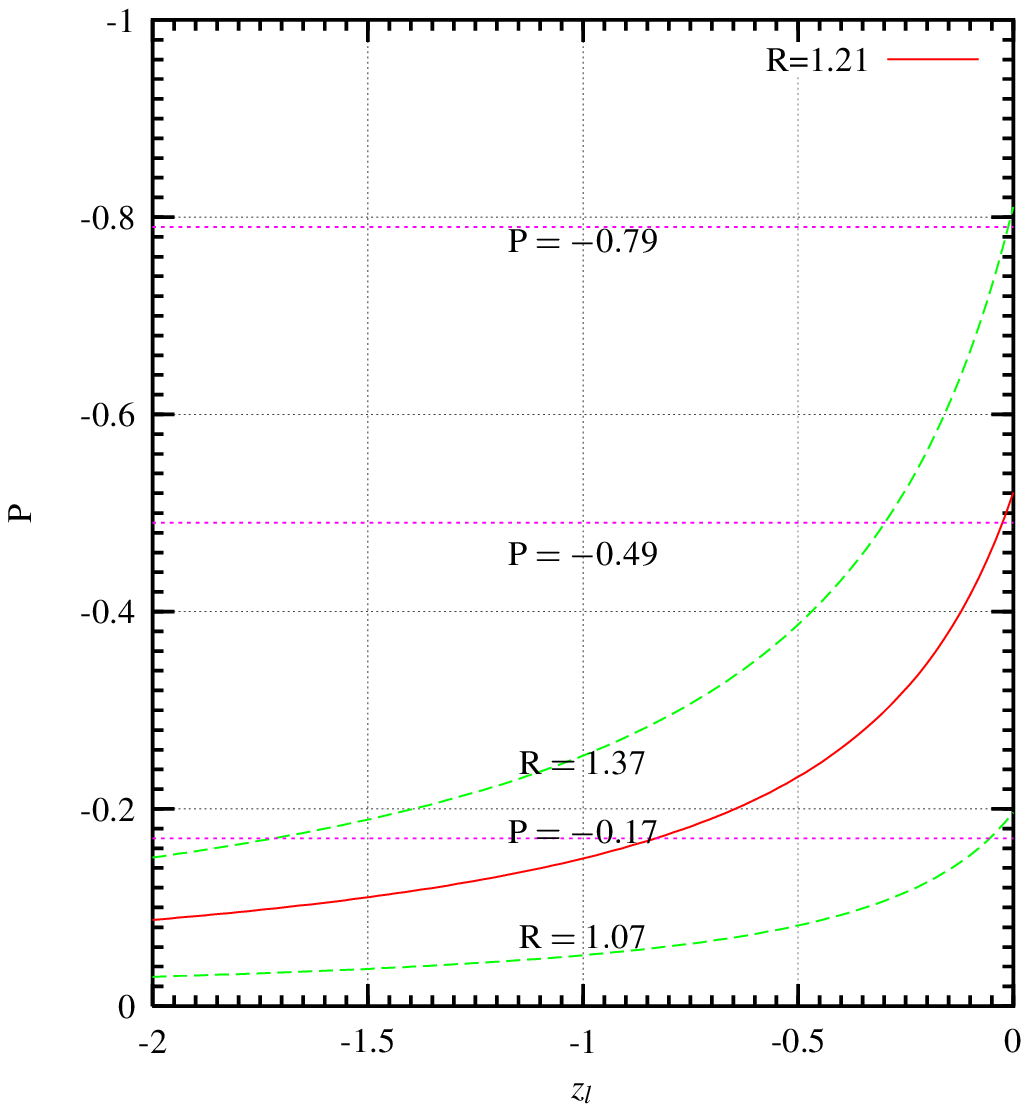}
\caption{Plot of $\rm P$ vs. $z_{\ell}$ is shown by employing the DELPHI ratio ${\rm R}=1.21^{+0.16}_{-0.14}$. The experimental polarization ${\rm P}=-0.49^{+0.32}_{-0.30}$ is also shown to indicate the allowed range of $z_{\ell}$.}
\end{figure*}
\newpage
\begin{figure*}
\includegraphics{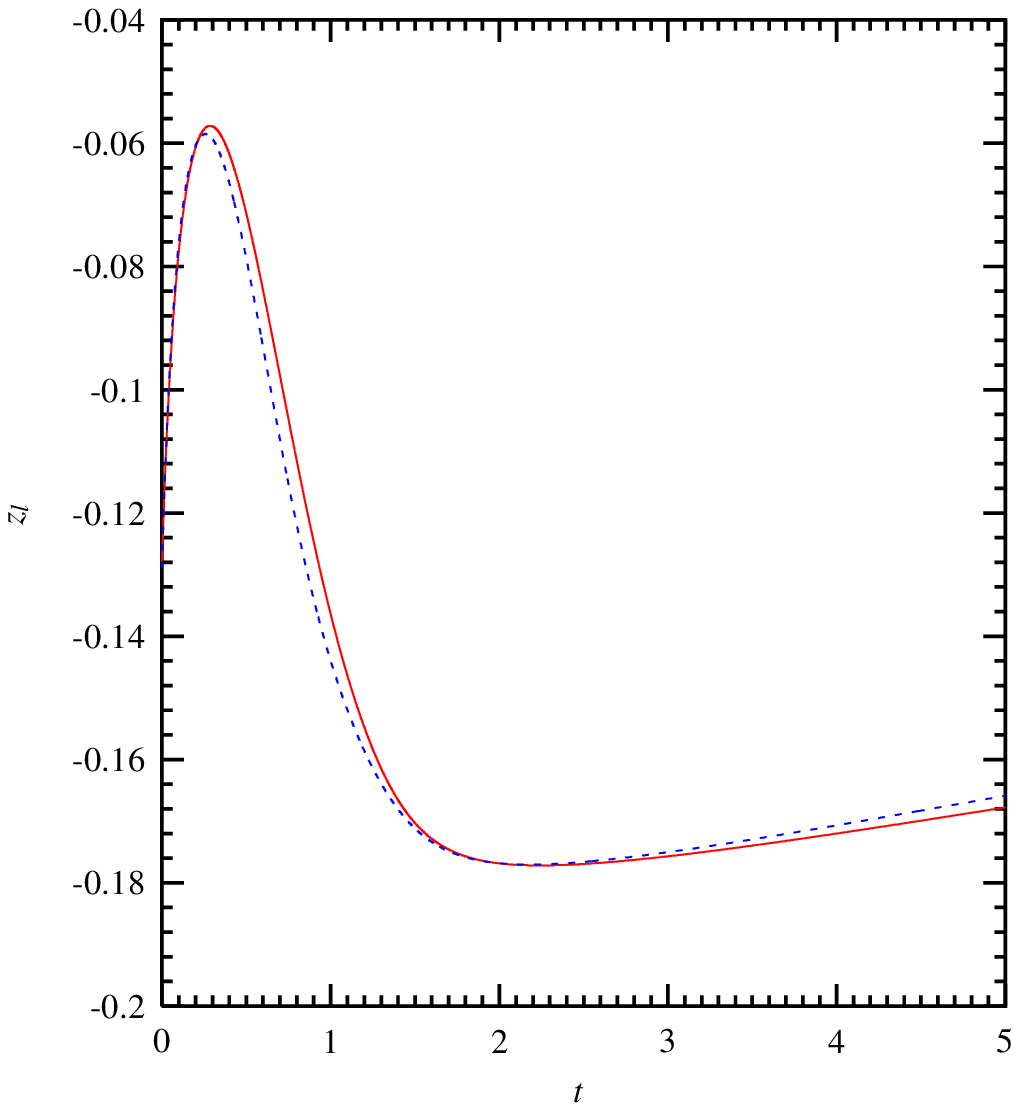}
\caption{\label{3} Plot of $z_\ell$ vs $t$. The modified quark model (solid line) and modified parton model (dashed line) are shown.}
\end{figure*}
\newpage

\begin{figure*}
\includegraphics{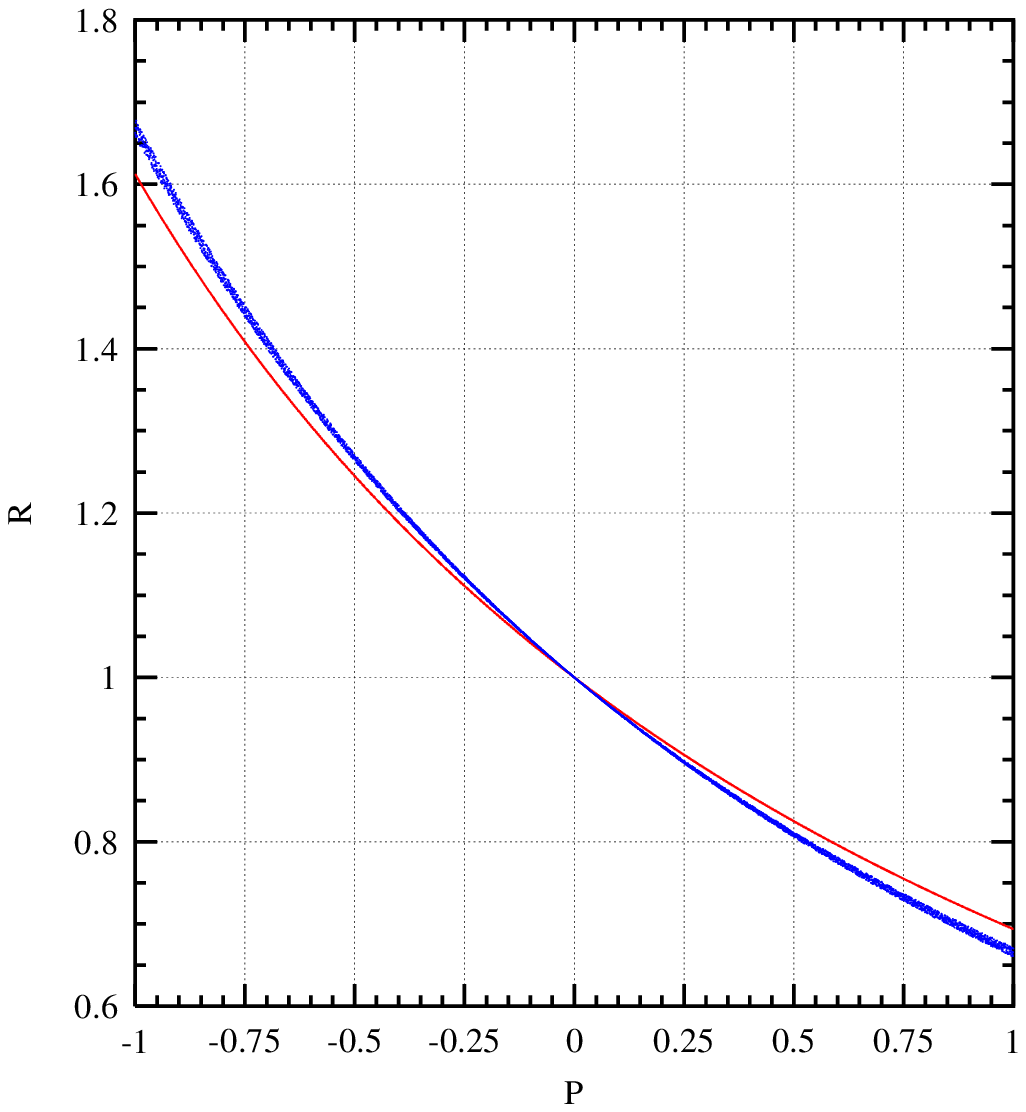}
\caption{\label{4} Plot of $\rm R$ vs. $\rm P$. The ALEPH and DELPHI Monte Carlo simulations (solid line) and the theoretical prediction (band line) are shown.}

\end{figure*}


\begin{thebibliography}{99}
\bibitem{ALEPH}
ALEPH Collaboration, \PLB {\bf B 365}, 437 (1996).
\bibitem{OPAL}
OPAL Collaboration, \PLB {\bf B 444}, 539 (1998).
\bibitem{DELPHI}
DELPHI Collaboration, \PLB {\bf B 474}, 205 (2000).
\bibitem{CLOSE}
F.E. Close, J.G. K\"oner, R.J.N. Philips and D.J. Summers, \JP {\bf G 8}, 1719(1992).
\bibitem{Mannel-Schular}
T.Mannel and G.A.Schular, \PLB {\bf B 279}, 194(1992).
\bibitem{KPTung}
J.G. K\"orner, A.Pilaftsis and M.Tung, \ZPC{\bf C 63}, 575(1994).
\bibitem{KRZ}
J.H. K$\ddot{u}$hn, A. Reiter and P.M. Zerwas, \NP {\bf B 272}, 560(1986).
\bibitem{JW}
S.Jadach and Z.Was, Acta. Phys. Pol. {\bf B 15}, 1151(1984).
\bibitem{FP}
A.F.Falk and M.E.Peskin, \PR {\bf D 49}, 3320(1994).
\bibitem{JK}
J. K\"orner, \NPPS {\bf 50}, 130(1994). 
\bibitem{BR}
G. Bonvicini and L. Randall, \PRL {\bf 73}, 392(1994).
\bibitem{LIYU1}
H-n. Li and H.L. Yu, \PR {\bf D53}, 4970(1996).
\end{thebibliography}
\end{document}